\documentclass[twoside,slac_one]{revtex4}
\usepackage{graphicx}
\usepackage{fancyhdr}
\usepackage{amsmath} 
\usepackage{bm}
\usepackage{amsxtra}
\usepackage{amssymb}
\usepackage{amsthm}
\usepackage{latexsym}
\usepackage{lscape}
\usepackage{color}
\newcommand{\hilight}[1]{\colorbox{yellow}{#1}}

\newcommand{\ltsimeq}{\raisebox{-0.6ex}{$\,\stackrel
        {\raisebox{-.2ex}{$\textstyle <$}}{\sim}\,$}}

\def \et {E_{T}}
\newcommand{\pte}{\ensuremath{p_T^e}}
\newcommand{\mt}{\ensuremath{m_T}}
\def  \met {\not\!\!\et }

\begin{document}

\title{Measurement of the W Boson Mass and Width at the D0 experiment}

\author{D. Boline}
\affiliation{Department of Physics and Astronomy, Stony Brook University, Stony Brook, NY, USA}

\begin{abstract}
I present a precise measurement of W boson mass measurement in electron decay channel using data collected by the D0 detector at the Fermilab Tevatron Collider. A binned likelihood fit method is used to extract the mass information from the transverse mass, the electron transverse momentum and missing transverse energy distributions. I also present a precise direct measurement of W boson width using the events with large transverse mass. The W mass result can be used to put stringent indirect limits on the Standard Model Higgs boson mass.
\end{abstract}

\maketitle

\thispagestyle{fancy}

\section{Introduction}

In 2009 the D\O\ experiment published precision measurements of both the W boson mass~\cite{Abazov:2009cp} and width~\cite{Abazov:2009vs}.
Both measurements make exclusive use of the decay channel $W\rightarrow e\nu$, due to the excellent energy precision of the D\O\ Calorimeter.
The uncertainty on our knowledge of $M_W$ is currently one of the strongest factor limiting our ability to constrain the mass of the standard model Higgs Boson.
The measurement of $\Gamma_W$ is important as a precision test of the Standard Model (SM), as within the SM the W boson width is related to the W boson mass through the expression given in Eq.~\ref{gweqn}.
\begin{equation}
\Gamma(W\rightarrow l\nu) = \frac{G_F}{\sqrt{2}} \frac{M_W^3}{6\pi} \left( 1 + \delta_{SM} \right)
\label{gweqn}
\end{equation}
Where $f_{QCD} = 3 ( 1 + \alpha_S(M_W^2)/\pi )$ to first order in $\alpha_S$ and $\delta_{SM} \approx 0.021 \pm 0.005$.

\subsection{Tevatron Collider}

We obtain our $W$ bosons from the Tevatron Collider at Fermi National Accelerator Laboratory (Fermilab).  The Tevatron collides protons and antiprotons at a beam energy of $980\mbox{ GeV}$ for a center-of-mass energy or $1.96\mbox{ TeV}$.
Both of these measurements make use of $1\mbox{ fb}^{-1}$ of data collected up to 2006, 
the Accelerator Division at Fermilab has done an excellent job of maximizing delivered luminosity to the D\O\ experiment, and as of September 30th we have $\approx 10\mbox{ fb}^{-1}$ of data available.

\subsection{D\O\ Detector}

\begin{figure}[h]
\centering
\includegraphics[width=80mm]{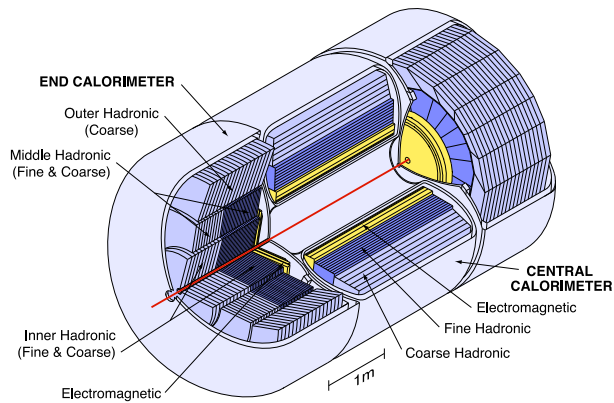}
\caption{D\O\ Calorimeter.} \label{d0calorimeter}
\end{figure}

Both of these measurements make extensive use of two components of the D\O\ Detector~\cite{Abazov:2005pn}, the Calorimeter to measure the electron energy and the hadronic recoil, and the Tracking Detector to identify electron tracks.
D\O\ uses a liquid Argon Calorimeter (Fig.~\ref{d0calorimeter}), with a Central Calorimeter (CC) extending to $|\eta|<1.1$, North and South End Cap (EC) Calorimeters from $1.5 < |\eta| \ltsimeq 4$, in addition there is a Scintillating Inter Cryostat Detector (ICD) between the CC and EC with $1.1<|\eta|<1.5$.
Both measurements principally use the CC calorimeter.
The CC is broken up longitudinally into an Electro-Magnetic (EM) portion and both a fine and course Hadronic (fHad,cHad) portion.  The EM portion is closer to the beam, and has a high precision for measuring electron energies.  The fHad portion contributes to the measurement of the recoil of the W boson, while the cHad is not used due to its larger noise.

\begin{figure}[h]
\centering
\includegraphics[width=80mm]{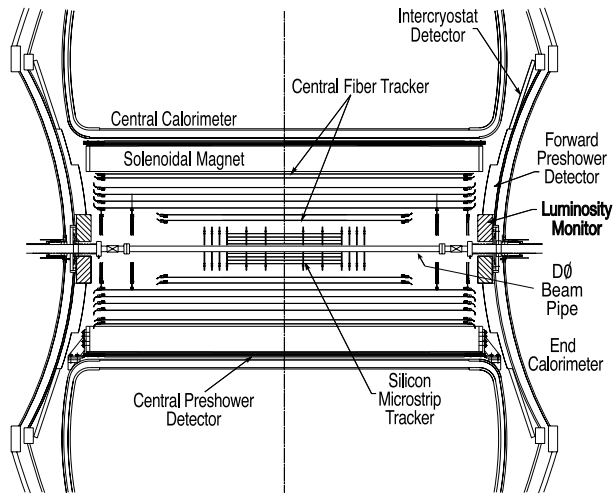}
\caption{D\O\ Tracking Detector.} \label{d0tracking}
\end{figure}

The Tracking Detector (Fig.~\ref{d0tracking}) is composed of two parts.
The Silicon Microstrip Tracker (SMT) sits closest to the beam, and provides a precise measurement of the z and $\phi$ of a given track, allowing for accurate vertexing.
The Central Fiber Tracker (CFT) is less precise in z, but offers high precision in $\phi$ and gives a precise measurement of the $p_T$ of high $p_T$ tracks.
Both tracking detectors sit within a Solenoid magnet with a $2\mbox{ T}$ field.

\section{Event Selection}

To select the $W$ boson sample, we require a single electron with:
\begin{itemize}
 \item $p_T^e > 25\mbox{ GeV}$
 \item $|\eta_e| < 1.05$
 \item Spatially matched track with hits in the SMT
\end{itemize}
We also require:
\begin{itemize}
 \item $\met > 25\mbox{ GeV}$
 \item $u_T < 15\mbox{ GeV}$, where $u_T$ is the recoil off the boson.
 \item $50 < m_T < 200\mbox{ GeV}$ where $m_T = \sqrt{ 2 p_T^e \met ( 1 - \cos{\theta} ) }$.
\end{itemize}
We obtain a sample of 499830 candidate events.

In addition to the $W$ boson sample used for the mass measurement we select a calibration sample composed of $Z$ bosons. For the $Z$ boson calibration sample, we require two electrons with the same selection criteria as the $W$ boson sample, $u_T < 15\mbox{ GeV}$, and in addition we require $70 < m_{ee} < 110\mbox{ GeV}$.  
We obtain a sample of 18725 candidate Z events.

\section{Backgrounds}

The three major backgrounds to this analysis come from $Z\rightarrow ee$, QCD multijet events, and $W\rightarrow\tau\nu_\tau\rightarrow e\nu_e\nu_\tau\nu_\tau$.
The $Z\rightarrow ee$ background arises from events where one electron passes the $W\rightarrow e\nu$ selection while the second electron falls into the the ICD region which this analysis doesn't use when calculating $\met$.
We estimate the size of the $Z\rightarrow ee$ background using events which pass the $W\rightarrow e\nu$ selection but also contain a track pointing towards the ICD region.  This background is estimated to make up $0.91\pm 0.01\%$ of the total $W\rightarrow e\nu$ sample.

The QCD multijet background arises when a jet fakes the signature of an electron.  
We estimate the size of the background using a sample without the requirement of a track match to the electron.
The size of this background is estimated to be $1.49\pm 0.03\%$ of the sample.

The $W\rightarrow \tau\nu$ background is irreducible, as it has both a high $p_T$ electron and large $\met$.
We estimate the size of this background using a sample of simulated events obtained using the PYTHIA~\cite{Sjostrand:2000wi} Monte Carlo event generator.  These events are estimated to comprise $1.60\pm 0.02\%$ or the sample.

\section{Monte Carlo Simulation}

A crucial part of both measurements is a precise theoretical model of both $W$ and $Z$ boson production.
To obtain this high level of precision we use the RESBOS event generator.  RESBOS combines a next-to-leading-order (NLO) calculation at high boson $p_T$ with a next-to-leading-log (NLL) resummation at low boson $p_T$, reproducing the observed $p_T$ spectrum in $Z\rightarrow ee$ events.
To simulate the effect of photon emission from final state electrons we use the PHOTOS subroutine.

Finally we use a detailed fast parametric Monte Carlo simulation (fastMC) to model the interaction of the particles produced in $W$ and $Z$ production with our detector.  This fastMC contains tuned models of both the electron shower and the hadronic recoil.

\section{Electron Energy Response}

\begin{figure}[h]
\centering
\includegraphics[width=80mm]{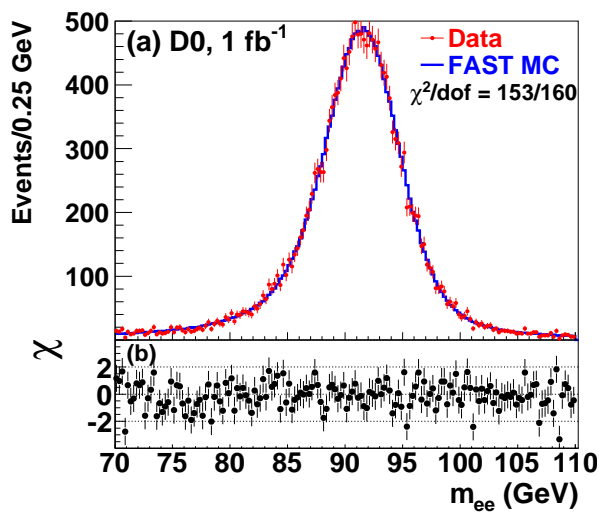}
\caption{Comparison between $Z$ peak in data and in fastMC after fitting electron energy scale and offset.} \label{fastMCdet}
\end{figure}

The most important component of our model of the electron shower is the energy response of the EM calorimeter.
The mass and width of the $Z$ boson were measured to a very high precision at the LEP Collider, and we use the values obtained there to calibrate our own detector.  This means that we are effectively measuring $M_W/M_Z$ and $\Gamma_W/M_Z$.
We model non-linearities in the energy response of our calorimeter through detailed GEANT~\cite{Brun:1978fy} simulations, then perform a fit to the remaining linear energy response.  The measured energy is related to the true energy by: $E^{measured} = \alpha E^{true} + \beta$, where $\alpha$ is the energy scale and $\beta$ is the offset.  
This relationship is put into our detailed fastMC, templates are formed with various values of $\alpha$ and $\beta$, and negative-log-likelihood fit is performed to find the values of $\alpha$ and $\beta$ most consistent with the observed $Z$ sample (Fig.~\ref{fastMCdet}).

\section{Hadronic Recoil Model}

We use a parametrized model to simulate the interaction of the hadronic recoil with our detector.
This parametrized model contains two parts, a hard component related to the $p_T$ of the boson, and a soft component coming from detector noise and underlying event.
The hard component is modeled using a special sample of $Z\rightarrow \nu\nu$ Monte Carlo events run through D\O's detailed GEANT simulation.
The soft component is broken into a zero bias (ZB) component, where there is no requirement for an interaction vertex in the event, and a minimum bias (MB) component, where exactly one interaction vertex is required.  Both the ZB and MB samples are obtained directly from data.
The final parametrized model is then tuned using the $Z\rightarrow ee$ data sample.

A second model of the hadronic recoil has also been obtained, using a library of recoil events from the $Z\rightarrow ee$ sample.  This model is used as a cross check and is found to be in good agreement with results obtained with the parameterized model.

\section{Mass Measurement}

\begin{figure}[h]
\centering
\includegraphics[width=80mm]{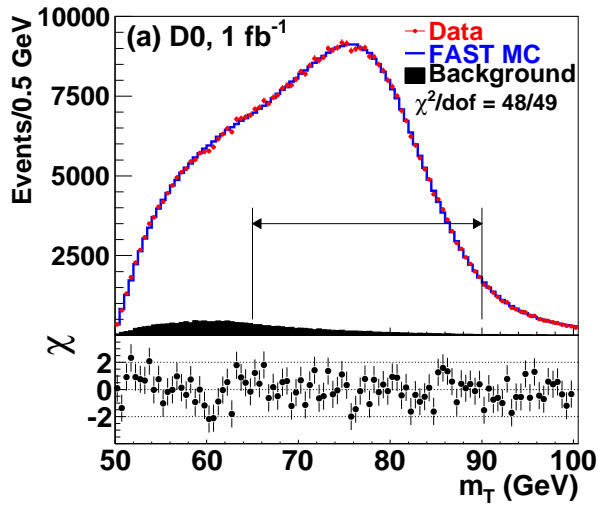}
\caption{$M_T$ distribution.} \label{mtdist}
\end{figure}

\begin{figure}[h]
\centering
\includegraphics[width=80mm]{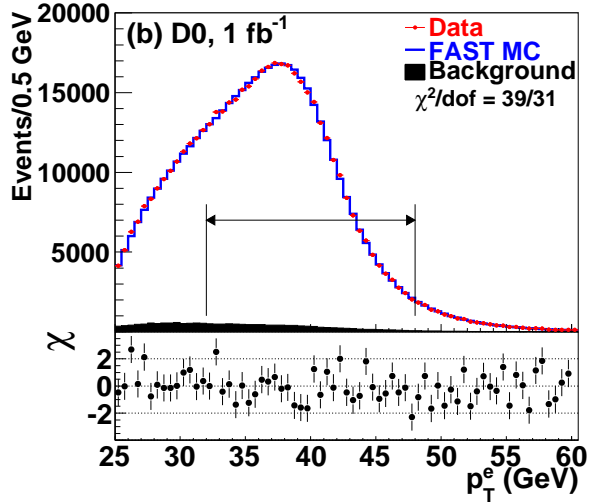}
\caption{$p_T$ distribution.} \label{ptdist}
\end{figure}

\begin{figure}[h]
\centering
\includegraphics[width=80mm]{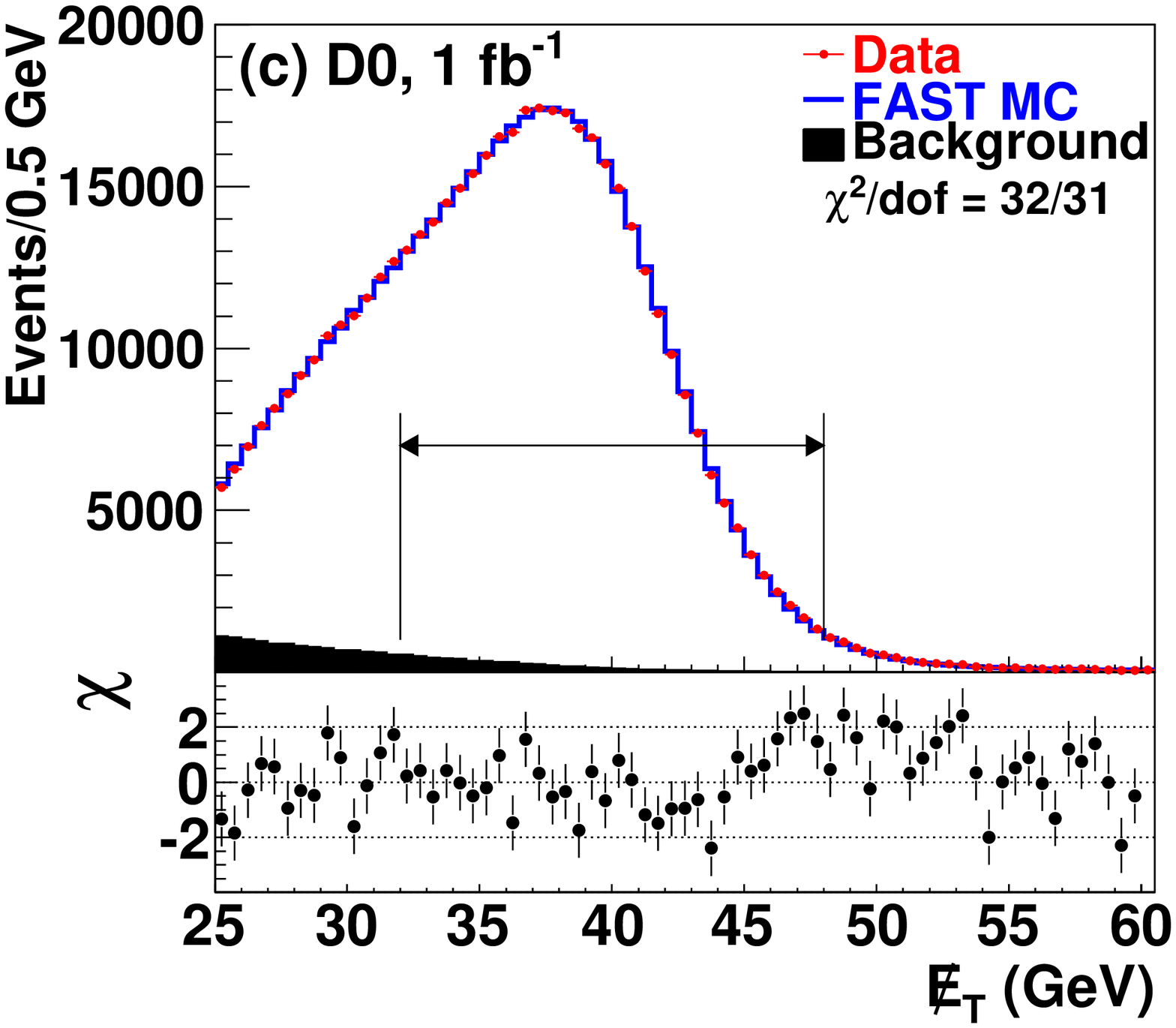}
\caption{$\met$ distribution.} \label{metdist}
\end{figure}

The measurement of the mass is performed using a binned likelihood fit comparing fastMC templates at set $M_W$'s to the data.  We repeat the measurement using the distribution of three observables: the transverse mass $m_T$ (Fig.~\ref{mtdist}), the transverse momentum of the electron $p_T^e$ (Fig.~\ref{ptdist}), and the missing transverse energy $\met$ (Fig.~\ref{metdist}).  In each case the fit range has been adjusted to minimize the total uncertainty.

\section{Systematic Uncertainties of the Mass Measurement}

The largest systematic uncertainty by far is the uncertainty on the electron energy calibration, and this is limited by the statistics of the $Z$ sample.
The largest theoretical uncertainty comes from the parton distribution functions (PDF's), it should be remarked that this uncertainty was obtained using the PYTHIA event generator, which doesn't accurately reproduce the boson $p_T$ spectrum.  The full list of uncertainties is given in Table~\ref{wmasssysttable}

\begin{table}[ht]
\begin{center}
\caption{Systematic Uncertainties of $M_W$ measurements.}
\begin{tabular}{l|ccc}
\hline\hline
  & \multicolumn{3}{c}{$\Delta M_W$~(MeV)} \\ \hline
   Source                          &$\mt$ & $\pte$ &  $\met$\\
  \hline \hline
  \hilight{Electron energy calibration       } & \hilight{34} &  \hilight{34} & \hilight{34} \\
  Electron resolution model         &  2 &   2 &  3 \\
  Electron shower modeling           &  4 &   6 &  7 \\
  Electron energy loss model        &  4 &   4 &  4 \\
  Hadronic recoil model             &  6 &  12 & 20 \\
  Electron efficiencies             &  5 &   6 &  5 \\
  Backgrounds                       &  2 &   5 &  4 \\ \hline
  Experimental Subtotal             & 35 &  37 & 41 \\ \hline
  \hilight{PDF                         } &  \hilight{10} &  \hilight{11} & \hilight{11} \\
  QED                          &  7 &   7 &  9 \\
  Boson $p_T$                  &  2 &   5 &  2 \\ \hline
  Production Subtotal          & 12 &  14 & 14 \\ \hline

  Total                        &  37 & 40 & 43 \\
\hline\hline
  \end{tabular}
\label{wmasssysttable}
\end{center}
\end{table}

\section{Width Measurement}

\begin{figure}[h]
\centering
\includegraphics[width=80mm]{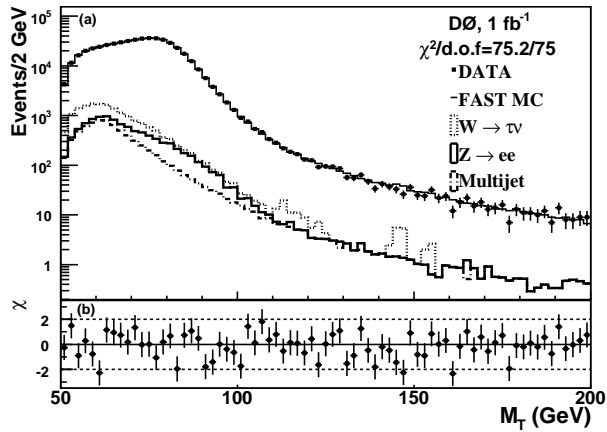}
\caption{$M_T$ distribution used in determination of W boson width.} \label{mtdistwidth}
\end{figure}

The W width measurement uses the $m_T$ distribution in the range $100 < m_T < 200\mbox{ GeV}$ (Fig.~\ref{mtdistwidth}).  As in the mass measurement templates are formed with varying values of $\Gamma_W$ and a fit is made to the data distribution.

\section{Systematic Uncertainties of the Width Measurement}

The systematic uncertainty for the width measurement is dominated by the recoil model (Table~\ref{gwsysttable}).

\begin{table}[ht]
\begin{center}
    \caption{Systematic uncertainties on the measurement of $\Gamma_W$.}
  \begin{tabular}{l|c} \hline \hline
   Source & $\Delta \Gamma_W$ (MeV) \\
  \hline
  \hilight{Electron energy scale            } & \hilight{33} \\
  Electron resolution model         & 10 \\
  \hilight{Recoil model }                     &  \hilight{41} \\
  Electron efficiencies             &  19  \\
  Backgrounds                       &  6 \\ 
  \hilight{PDF}                               &  \hilight{20} \\
  Electroweak radiative corrections          &  7 \\
  Boson $p_T$                       &  1 \\ 
  $M_W$                             &  5 \\ \hline
  Total Systematic                  &  61 \\ \hline \hline 
  \end{tabular}
\label{gwsysttable}
\end{center}
\end{table}

\section{Results}

\begin{figure}[h]
\centering
\includegraphics[trim= 80 30 47 51,clip=true,width=80mm]{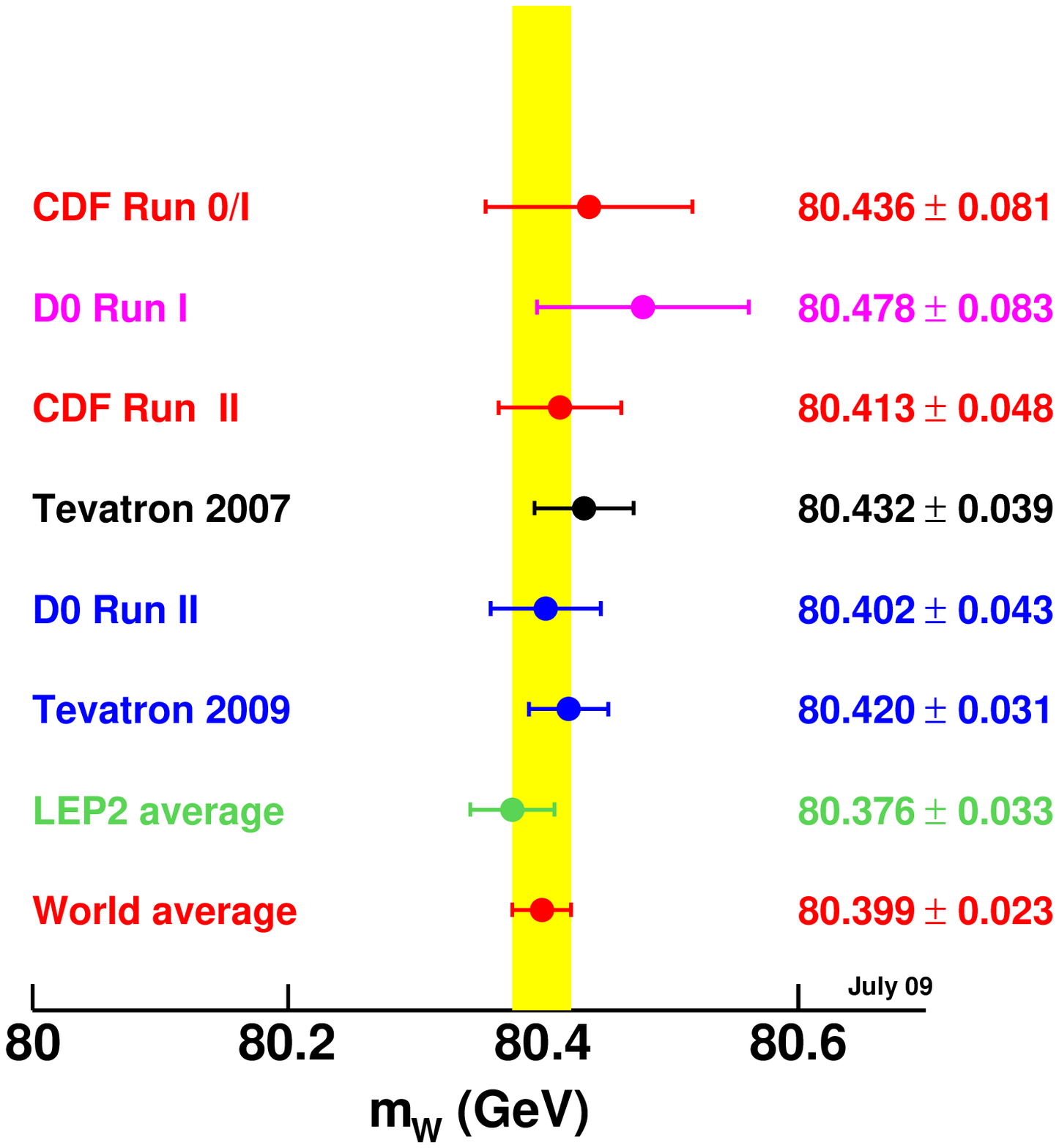}
\caption{$m_W$ measurement at D\O\ compared with other experiments and SM theory.} \label{mwcomp}
\end{figure}

\begin{figure}[h]
\centering
\includegraphics[trim= 15 0 0 0 ,clip=true,width=80mm]{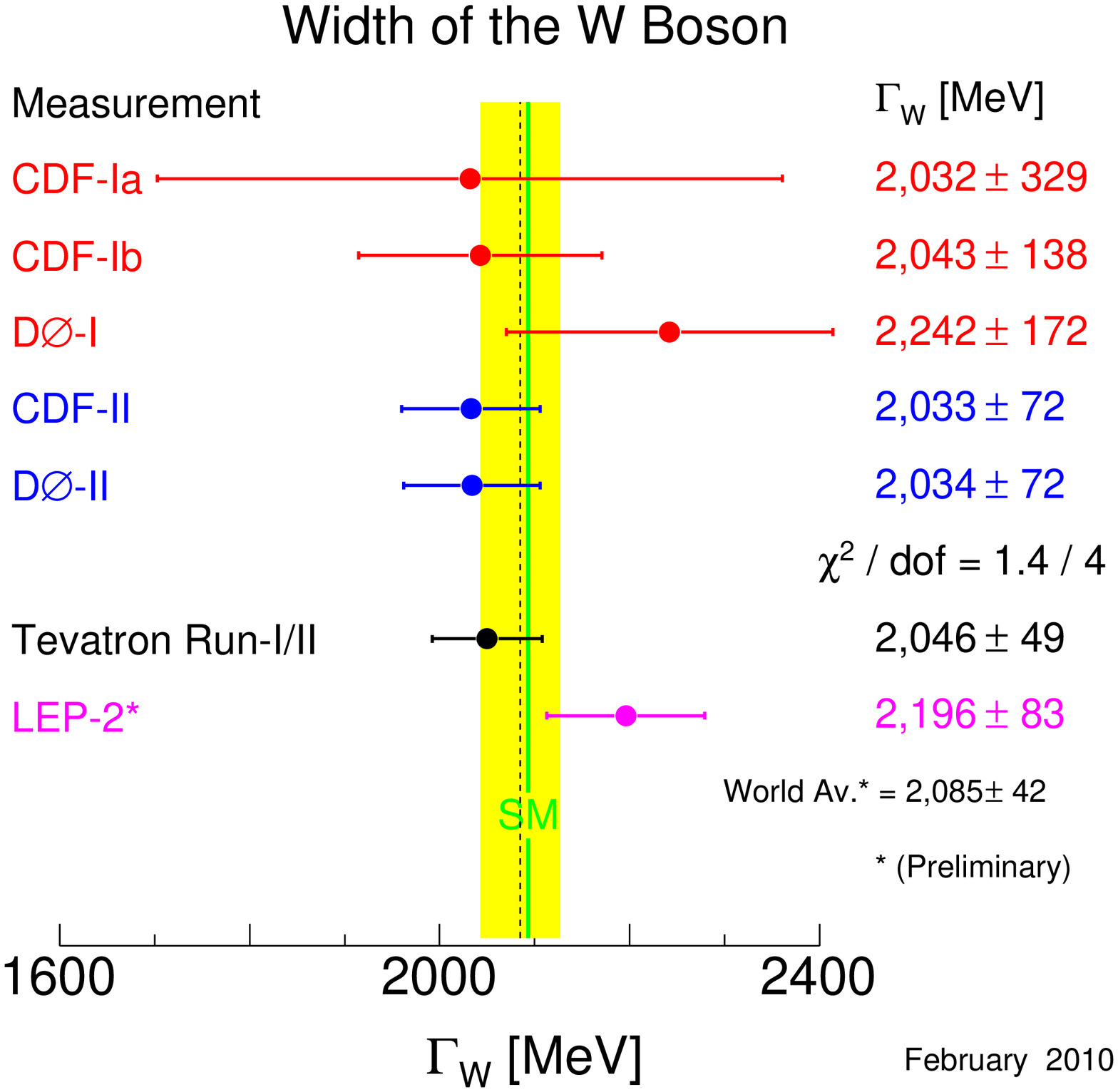}
\caption{$\Gamma_W$ measurement at D\O\ compared with other experiments and SM theory.} \label{gwcomp}
\end{figure}

The results of our measurements are given in Eq.~\ref{mw_value} and ~\ref{gw_value}.  Comparisons to the CDF and LEP measurements, along with the combination are given in Fig.'s~\ref{mwcomp} and ~\ref{gwcomp}, for the combination the measured width is adjusted to use the measured mass and the measured mass is adjusted to use the measured width~\cite{:2009nu}~\cite{:2010in}, thus the inconsistency between Eq.'s ~\ref{mw_value} - ~\ref{gw_value} and Fig.'s ~\ref{mwcomp} - ~\ref{gwcomp} .
\begin{eqnarray}
 m_W &=& 80.401 \pm 0.021\mbox{ (stat) } \pm 0.038\mbox{ (syst) GeV} \nonumber \\
     &=& 80.401 \pm 0.043\mbox{ GeV} \label{mw_value} \\
 \Gamma_W &=& 2.028 \pm 0.039\mbox{ (stat) } \pm 0.061\mbox{ (syst) GeV} \nonumber \\
          &=& 2.028 \pm 0.072\mbox{ GeV} \label{gw_value}
\end{eqnarray}

\section{Conclusion}

The D\O\ $m_W$ measurement is currently the most precise single experiment result.
We have to date collected an order-of-magnitude more data than the results presented.
We expect a substantial improvement in our precision.
With $10\mbox{ fb}^{-1}$ we anticipate a electron energy scale uncertainty of $\approx 15\mbox{ MeV}$ and a total systematic uncertainty of $\approx 25\mbox{ Mev}$.

\bigskip 

\end{document}